\documentclass[a4,aps,amsmath]{revtex4}
\usepackage{graphicx}
\usepackage{color}
\newcommand{\be}{\begin{equation}}
\newcommand{\ee}{\end{equation}}
\newcommand{\ba}{\begin{array}}
\newcommand{\ea}{\end{array}}
\newcommand{\bea}{\begin{eqnarray}}
\newcommand{\eea}{\end{eqnarray}}
\newcommand{\bdm}{\begin{displaymath}}
\newcommand{\edm}{\end{displaymath}}

\begin{document}

\title{Effect of Coordination Number on Nonequilibrium Critical Point}

\author{Diana Thongjaomayum and Prabodh Shukla}
\affiliation{%
Department of Physics, \\ North Eastern Hill University, \\ 
Shillong 793 022, India}%


\begin{abstract} 

We study the nonequilibrium critical point of the zero-temperature 
random-field Ising model on a triangular lattice and compare it with 
known results on honeycomb, square, and simple cubic lattices. We 
suggest that the coordination number of the lattice rather than its 
dimension plays the key role in determining the universality class of the 
nonequilibrium critical behavior. This is discussed in the context of 
numerical evidence that equilibrium and nonequilibrium critical points 
of the zero-temperature random-field Ising model belong to the same 
universality class. The physics of this curious result is not fully 
understood.

\end{abstract}

\maketitle

\section{Introduction}

Systems with quenched disorder tend to respond to a smoothly varying 
force by way of avalanches in their structure ~\cite{sethna1, 
bertotti,durin,fisher1,miguel}. Generally, the avalanches are 
microscopic and therefore the response is macroscopically smooth. 
However, there are several instances in nature where the response 
exhibits an abrupt and catastrophic jump discontinuity: snow falls 
gently on a mountain for days before tons of snow may move abruptly 
in an avalanche. This cannot be attributed to the last snowflake 
before the avalanche. Similarly, landslides may occur abruptly during 
periods of incessant rain. Earthquakes are sporadic and sudden 
effects of the tectonic plates pressing against each other all the 
time. A more familiar and repeatable example from a physics 
laboratory is the Barkhausen noise on magnetization 
curves~\cite{stoner}. The theory of the Barkhausen noise has been 
studied extensively in the framework of the zero-temperature 
nonequilibrium random-field Ising model of a ferromagnet on a 
lattice~\cite{sethna2}. The disorder is normally modeled by an 
on-site Gaussian random-field with average value zero and standard 
deviation $\sigma$. Numerical simulations on $d$-dimensional lattices 
($d>1$) reveal that there is a critical value $\sigma_c$ such that 
for $\sigma>\sigma_c$, the response of the system, i.e., the 
magnetization $m(h,\sigma)$ in an applied field $h$, is 
macroscopically smooth over the entire range of the applied field 
$-\infty< h<\infty$. For $\sigma < \sigma_c$, $m(h,\sigma)$ has a 
jump discontinuity at $h=h_c(\sigma)$. The size of the jump as well 
as $|h_c(\sigma)|$ decreases as $\sigma \rightarrow \sigma_c$. The 
point $\{\sigma_c,h_c(\sigma_c)\}$ is a nonequilibrium critical point 
exhibiting anomalous scale-invariant fluctuations and universality 
reminiscent of the equilibrium critical point phenomena.

The similarity between the nonequilibrium and equilibrium critical 
behavior has a reason, but one that is not fully understood. The 
equilibrium critical point in a system with quenched disorder is 
controlled by a stable zero-temperature fixed point~\cite{fisher2}. 
Therefore, it is not entirely surprising to observe critical behavior 
at zero temperature by varying the parameter $\sigma$. Let us 
consider zero-temperature equilibrium magnetization curves for 
different values of $\sigma$. We may expect smooth trajectories for 
$\sigma>\sigma_c$, jump discontinuities for $\sigma<\sigma_c$, and a 
vanishing jump discontinuity at $\sigma=\sigma_c$ just as in the case 
of the nonequilibrium magnetization curves. The only difference will 
be that the equilibrium case would show no hysteresis and therefore 
all singularities will occur at $h=0$. The nonequilibrium response 
will show hysteresis and the singularities for $\sigma\le\sigma_c$ 
will occur on the lower and upper halves of the hysteresis loop in a 
symmetrical fashion. One may ask if the noise on the equilibrium 
magnetization curves and the anomalous fluctuations at the 
equilibrium critical points have the same character as their 
nonequilibrium counterparts. Surprisingly, the answer to this 
question is that they do in the framework of the zero-temperature 
random-field Ising model. Numerical studies of the model provide 
strong evidence that the disorder-induced critical points in 
equilibrium as well as the nonequilibrium case belong to the same 
universality class~\cite{liu}. This serendipity may provide a way to 
infer nonequilibrium properties of a system from its equilibrium 
properties. It is intriguing. As a system is driven by an applied 
field from $h=-\infty$ to $\infty$, the nonequilibrium trajectory of 
the system comprises a sequence of metastable states. The equilibrium 
trajectory goes through the states of global minima of the energy. 
What is the physics that puts the two cases in the same universality 
class?

We focus on a similar but smaller question. It is known that the 
universality class of an equilibrium critical point is determined by 
the dimensionality of the lattice and not by the kind of Bravais 
lattice it is. The reason for this is well understood. The 
correlation length diverges at the critical point and therefore the 
short-range structure of the lattice is irrelevant to the critical 
behavior. In contrast, the existence of a nonequilibrium critical 
point, let alone its universality class, appears to be determined by 
the coordination number of the lattice rather than its 
dimensionality. An exact calculation on the Bethe lattice of an 
arbitrary coordination number $z$ shows that a nonequilibrium 
critical point exists only if $z>3$ ~\cite{dhar}. Numerical 
simulations suggest that the significance of this result goes beyond 
the Bethe lattice; periodic lattices with $z=3$ in $d=2,3$ do not 
possess a nonequilibrium critical point ~\cite{sabhapandit}. Although 
these results have been reported quite some time ago, their 
significance does not appear to be widely recognized. For example, 
numerical simulations have often failed to settle the question if 
there is a nonequilibrium critical point on the square lattice 
($z=4$)~\cite{bertotti}. Recent results indicate that it is 
there~\cite{spasojevic}. This has been taken to indicate that the 
lower critical dimension for nonequilibrium critical behavior is 
equal to 2. However, it has been shown earlier that there is no 
nonequilibrium phase transition on a honeycomb 
lattice~\cite{sabhapandit}. Indeed, there is good evidence that the 
nonequilibrium critical behavior is controlled by a lower critical 
coordination number ($z=4$) rather than a lower critical dimension 
($d=2$).

In this paper we study the nonequilibrium critical point on a 
triangular lattice ($z=6$) reiterating the importance of the 
coordination number rather than the dimensionality of the lattice. We 
show the existence of a critical point and estimate the critical 
exponent $\nu$. Within numerical errors we find $\nu$ to be 
tantalizingly close to the value reported for the simple cubic 
lattice. The simple cubic and the triangular lattices have the same 
coordination number ($z=6$) and it is possible that the values of the 
nonequilibrium critical exponents depend on $z$ just as their 
existence depends on $z$. Indeed, it may be useful to study the 
triangular lattice with gradual dilution of one of the three 
sublattices comprising it. This way, one can go continuously from a 
triangular ($z=6$) to a honeycomb $(z=3)$ lattice and study the 
effect of $z$ on the critical behavior. However, this is beyond the 
scope of the present paper.

\section{The Model and Simulations on a Triangular Lattice}

The random-field Ising model is characterized by the Hamiltonian, \be 
H=-J\sum_{i,j} s_is_j -\sum_{i}h_is_i-h\sum_is_i, \ee where 
$\{s_i=\pm1\}$ are Ising spins on a triangular lattice and $\{h_i\}$ 
are identically distributed independent random fields drawn from a 
Gaussian distribution with mean zero and standard deviation $\sigma$. 
Periodic boundary conditions are imposed. Here $J$ is the 
ferromagnetic interaction between nearest neighbors and $h$ is an 
external field that is varied adiabatically from $-\infty$ to 
$\infty$. A stable state of the system at $h$ has each spin aligned 
along the net field at its site. As $h$ is ramped up, numerous 
instabilities occur where a spin flips up and causes neighboring 
spins to flip up in an avalanche. When this happens, $h$ is kept 
constant during the avalanche and then increased again until the next 
avalanche. The curve $m(h,\sigma)$ is the locus of the magnetization 
of locally stable states along increasing $h$ for a random-field 
distribution characterized by standard deviation $\sigma$.

In contrast to the case of the square lattice~\cite{spasojevic}, it 
takes a rather modest effort to see that the curve $m(h,\sigma)$ 
makes a transition between a discontinuous and a continuous form as 
$\sigma$ is increased. Fig. 1 shows $m(h,\sigma)$ in increasing $h$ 
for $\sigma$=1 (red triangles), $\sigma$=1.275 (orange squares), and 
$\sigma$=2 (blue continuous line) on a $1000\times1000$ triangular 
lattice. Only the data in the range $1.3\le h \le 2$ are shown. The 
curve for $\sigma=1$ shows a jump in the magnetization at 
$h\approx1.9$, but the curve for $\sigma=2$ is smooth. This suggests 
that a transition occurs at a critical value $\sigma_c$ 
($1<\sigma_c<2$) as $\sigma$ is increased. However, it is difficult 
to locate the exact $\sigma_c$. The difficulty is illustrated by the 
$m(h,\sigma)$ curve for $\sigma=1.275$, which is close to the 
critical value $\sigma_c$. Ideally, we would like to see a single 
jump in an otherwise smooth curve, and the size of the jump 
approaching zero as $\sigma\rightarrow\sigma_c$ from below. However, 
the critical point $\{\sigma_c,h_c\}$ is characterized by anomalously 
large fluctuations. Therefore, the critical curve in a typical 
simulation is punctuated by several jumps of different sizes. 
Increasing the system size does not alleviate this difficulty because 
the critical fluctuations also increase in proportion. We will return 
to this point in the following paragraph. For now we note another 
point of caution even in the case $\sigma << \sigma_c$ where a large 
jump in $m(h,\sigma)$ is rather obvious. In the limit $\sigma 
\rightarrow 0$, the initial state of the system ($h=-\infty$, all 
spins down) has an instability such that the first spin to flip up in 
increasing $h$ causes all the spins in the system to flip up. This is 
easily understood. The first spin flips up at $h=6J-h_{max}$, where 
$h_{max}$ is the maximum random field on an $L \times L$ lattice, 
$h_{max}^2\approx \sigma^2 \ln{2\pi \sigma^2/L^4}$. After the first 
spin flips up, the effective field on its neighbors becomes $4J-h_i$, 
which is positive with probability unity in the limit 
$\sigma\rightarrow 0$, and so all the neighbors flip up. Indeed, this 
causes an infinite avalanche leading to a state with all spins up. It 
is important to distinguish this instability from a genuine 
disorder-driven discontinuity that may occur for larger values of 
$\sigma$~\cite{sabhapandit}.

It is relatively easy to spot a large first-order discontinuity in 
the magnetization curve $m(h,\sigma)$. However, it is not as 
straightforward as it may look at first sight. In simulations as well 
as in experiments, it is difficult to distinguish between a truly 
discontinuous curve and one that may be smooth but steeply rising. We 
have to employ a method that takes into account the nature of 
fluctuations underlying the phase transition. One of the methods used 
in the literature is that of the Binder cumulant~\cite{binder} 
calculated from the averages of the square and the fourth power of 
the magnetization. It has been used for estimating the critical 
temperature of Ising models and distinguishing between first-order 
and second-order phase transitions in models without quenched 
disorder and applied field. For the present problem, we use a method 
that counts all the avalanches of size $s$ as the system is driven 
from $h=-\infty$ to $\infty$. Let $P(s,\sigma)$ be the probability of 
an avalanche of size $s$, where $\sigma$ is the standard deviation of 
the random-field distribution. In general, $P(s,\sigma)$ is a product 
of an algebraically decreasing part and an exponentially decreasing 
part with a cutoff $s_0$ that sets the scale of the avalanches. At a 
critical point we have $s_0=\infty$ and therefore the avalanches 
become scale invariant.

The distribution $P(s,\sigma)$ has a different form depending upon 
whether $m(h,\sigma)$ is continuous or discontinuous ~\cite{farrow}. 
The idea is illustrated by Fig. 2, which shows the probability 
$P(s,\sigma)$ of an avalanche of size $s$ for $\sigma$=1.25 (red 
triangles), 1.63 (pink squares), and 2 (blue), respectively. The 
$m(h,\sigma)$ curve for $\sigma=1.25$ has a jump discontinuity. The 
avalanches near the discontinuity are very large and may span the 
entire system. Away from the discontinuity the avalanches are small 
and $P(s,\sigma)$ decreases exponentially with $s$. Thus the defining 
trend of the avalanche distribution (red triangles) along a 
discontinuous magnetization curve is an algebraically decreasing 
$P(s,\sigma)$ followed by a peak at $s \approx L \times L$. In 
contrast to this, the avalanches along a noncritical continuous 
magnetization curve, e.g., the $m(h,\sigma)$ curve for $\sigma=2$ in 
Fig. 1, are exponentially small with a cutoff much smaller than 
$L\times L$. This is reflected in the corresponding curve (blue 
circles) in Fig. 2 by an initial algebraic decrease of $P(s,\sigma)$ 
followed by a more rapid decrease characteristic of the cutoff. Thus 
the distribution of avalanche sizes along the magnetization curve 
provides us with a method to distinguish between a smooth 
$m(h,\sigma)$ and one with a discontinuity. However, our goal is to 
identify a critical $m(h,\sigma)$ curve where the discontinuity just 
vanishes, i.e., to determine $\sigma_c$. This is evidently a 
difficult task. At $\sigma_c$, we may expect $\ln{P(s,\sigma)}$ to 
vary linearly with $\ln{s}$ in the entire range $1<s<L\times L$ with 
the peak at $s\approx L\times L$ just vanishing. It is difficult to 
implement this criterion strictly within a reasonable computational 
effort because avalanche distributions for different $\sigma$ have to 
be obtained and compared with each other. We have tried to meet this 
criterion within a reasonable error to the second decimal place in 
$\sigma_c$ as illustrated by the pink (squares) curve in Fig. 2 for 
$\sigma=1.63$; $\sigma_c=1.63\pm0.01$ is our best estimate for the 
critical point on the $100\times 100$ triangular lattice. This 
estimate has been obtained from 50000 independent realizations of the 
random-field distribution and took nearly a day of CPU time on our 
computer. We have often used binned data along with the unbinned data 
to find the best estimate for $\sigma_c$. As an illustration, Fig. 3 
shows the binned data for avalanches on a $200\times200$ lattice. The 
possible range of an avalanche lies between a single spin flip and 
$4\times10^4$ spin flips. This range is divided into 40 linear bins 
and the weight of each bin is represented by a point on the curve. 
See the caption on Fig. 3 for more details. We have also analyzed the 
data shown in Fig. 2 using logarithmic binning. The result is shown 
in Figure (4). As may be expected, the fat tails of the distributions 
shown in Fig. 2 are replaced by more clearly defined curves.

\section{Finite Size Effects}

Following the procedure outlined above, we have determined $\sigma_c(L)$ 
for lattices of linear size $L=100-400$. The results are 
presented in Table I.

\begin{table}[h]
\caption{$\sigma_c(L)$ for lattices of linear size $L=100-400$. }
\centering
\begin{tabular}{c c}
\hline \hline
$L$ & $\sigma_c$(L) \\ [0.5ex]
100 &	1.63 $\pm$ 0.01 \\
125 &	1.58 $\pm$ 0.01 \\
150 &	1.545 $\pm$ 0.005 \\
175 &	1.52 $\pm$ 0.01 \\
200 &	1.50 $\pm$ 0.01 \\
250 &	1.47 $\pm$ 0.01 \\
300 &	1.45 $\pm$ 0.01 \\
400 &	1.42 $\pm$ 0.01 \\ [1ex]
\end{tabular}
\end{table}

As $\sigma \rightarrow \sigma_c$ from below, the size of the avalanche 
diverges with the exponent $\nu$, i.e., $s \sim (\sigma 
-\sigma_c)^{-\nu}$. On a finite lattice, the largest avalanche is 
limited by the size of the lattice. Thus we define a 
lattice-dependent 
critical value $\sigma_c(L)$ by the equation

\be L^{-\frac{1}{\nu}}=\frac{\sigma_c(L)-\sigma_c}{\sigma_c} \mbox{   
or   } -\frac{1}{\nu} \log_{10} L = \log_{10} \left[ 
\frac{\sigma_c(L)}{\sigma_c} 
-1\right]\ee

We determine $\sigma_c$ by requiring the data in Table I to fit a 
straight line. The best fit to the straight line is shown in Fig. 5. 
The slope of the line gives $1/\nu=0.62$, or $\nu=1.6 \pm 0.2$. The 
data shown in Table I are based on linear binning. They change 
slightly if logarithmic binning is used; we get $\sigma_c(L)=1.525 
\pm 0.005$ for $L=175$, $\sigma_c(L)=1.41 \pm 0.01$ for $L=400$, and 
estimates of $\sigma_c(L)$ for other values of $L$ remain unchanged. 
The quality of the best straight line fit to the changed data is 
slightly poorer as compared to the one shown in Fig. 5, but it yields 
$\nu=1.5 \pm 0.2$.

Although it is a good practice to fit the data with a minimum number of 
adjustable parameters, we also tried the following form with an 
additional parameter $a$:

\be L^{-\frac{1}{\nu}}=\frac{\sigma_c(L)-\sigma_c^\prime}{a}\ee

We find that the best straight line fit to Eq. (3) is obtained when 
$\sigma_c^\prime= \sigma_c$ obtained from fitting the data to Eq. (2). 
If we set $\sigma_c^\prime=0$, we are not able to fit the data to a 
straight line. This shows that we must have $\sigma_c^\prime > 0$. 
Figure 6 shows the data and the nearest straight line fit to it for 
$\sigma_c^\prime=0$ and $a=1$. The slope of the straight line 
corresponds to $\nu=10.2$ approximately. The role of the parameter $a$ 
is only to shift the curve along the $y$ axis. It does not affect the 
slope of the straight line that best fits the data.

The existence of a critical point is expected to be accompanied by 
scaling of thermodynamic functions in its vicinity. Thus the existence 
of a critical value $\sigma_c >0$ means that a quantity such as 
$P(s,\sigma)$, which is in general a function of two independent 
variables $s$ and $\sigma$, must become a function of a single 
variable, 
say, $s|\frac{\sigma_c-\sigma}{\sigma}|^p$ as $\sigma \rightarrow 
\sigma_c$ for some value of the exponent $p$. Let us define

\be r=\frac{\sigma_c(L)-\sigma}{\sigma}.\ee

The scaling hypothesis requires that as $\sigma \rightarrow 
\sigma_c(L)$, the plots of $s^q P(s,\sigma)$ vs $s|r|^p$ for a fixed 
lattice of size $L \times L$ and different values of $\sigma$ should 
collapse on a single curve for suitable choices of the exponents $p$ 
and $q$. Figure 7 shows such a collapse for a $200 \times 200$ 
lattice. The collapsed curves are distinct for $r<0$ ($q=2$ and 
$p=0.66$) and $r>0$ ($q=2$ and $p=0.25$). The exponents $p$ and $q$ 
may be related to standard critical point exponents, e.g., 
$q=\tau+\sigma\beta\delta$ ~\cite{spasojevic,perkovic2}. The family 
of collapsed curves for different thermodynamic functions can be used 
to determine standard critical exponents, but in the present case we 
do not have sufficient data to take this approach. We are content to 
note that the avalanche size distributions show a reasonable collapse 
in a rather wide region around the critical point.

\section{Discussion}

We have shown that there is a critical point in the behavior of the 
zero-temperature random-field Ising model on a triangular lattice. This 
result assumes significance in the context of a long-standing 
speculation as to whether there is a nonequilibrium critical point in 
two dimensions. Generally, the hysteretic behavior of the 
random-field 
Ising model on a square lattice has been taken to characterize the 
behavior of the model in two dimensions. However, the coordination 
number $z$ of the lattice seems to be a key parameter. Two-dimensional 
lattices with $z=3$ (honeycomb structure) do not have a critical 
point~\cite{sabhapandit}. Studies on a square lattice ($z=4$), were 
initially inconclusive but more recent studies suggest that it has a 
critical point~\cite{spasojevic}. The existence of a critical point on a 
triangular lattice ($z=6$) can be verified with a rather modest effort 
as shown here. These results suggest that a lower critical coordination 
number has a greater significance for determining critical avalanches 
than a lower critical dimension. Although this goes against the spirit 
of the renormalization group theory that the short-range structure of 
the lattice should become irrelevant at the divergence of the 
correlation length,  some reflection shows that it is reasonable in 
the context of avalanches. It is reasonable that the coordination number 
of the lattice should determine how far an avalanche can propagate from 
its point of origin. Therefore, a minimum coordination number must be 
necessary for the divergence of avalanches irrespective of the 
dimensionality of the lattice.

If the existence of the nonequilibrium critical point depends on a 
lower critical coordination number rather than a lower critical 
dimension of the lattice, then it is natural to ask if the critical 
exponents depend upon $z$ as well. An exact solution of the 
random-field Ising model on a Bethe lattice of coordination number 
$z$ shows that the exponents are independent of $z$ as long as $z \ge 
4$. Numerical results on periodic lattices do not give such a clear 
indication. Let us focus on the critical exponent $\nu$. The 
uncertainty in the numerical determination of this exponent is rather 
large, although it is of central importance conceptually. The best 
estimates are $\nu=5.15\pm0.20$ on a square ($z=4$) 
lattice~\cite{spasojevic}, $\nu=1.4\pm0.2$ on a simple cubic ($z=6$) 
lattice~\cite{perkovic}, and the present result $\nu=1.6\pm0.2$ on a 
triangular ($z=6$) lattice. The closeness of $\nu$ on simple cubic 
and triangular lattice is interesting in view of the fact that the 
coordination number of both lattices is the same.

\begin{figure}[htb] 
\includegraphics[width=.5\textwidth,angle=0]{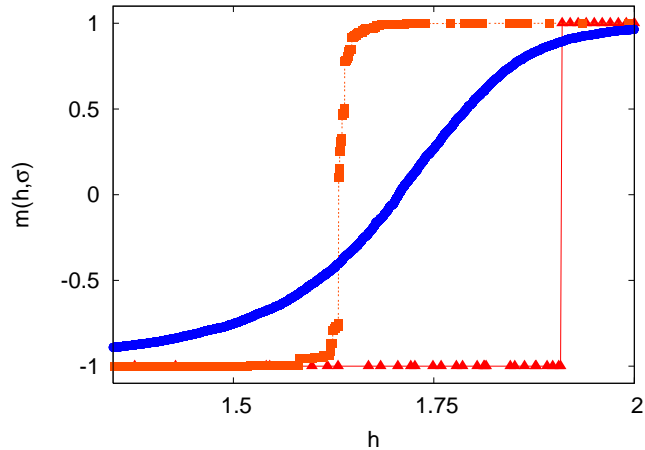} 
\caption{(Color online) Magnetization curves in increasing field $h$ 
for $\sigma=1$ (red discontinuous curve comprising triangles), 
$\sigma=2$ (blue continuous curve), and at an intermediate value 
$\sigma=1.275$ (orange squares with several jumps).} \label{fig1} 
\end{figure}

\begin{figure}[htb] 
\includegraphics[width=.5\textwidth,angle=0]{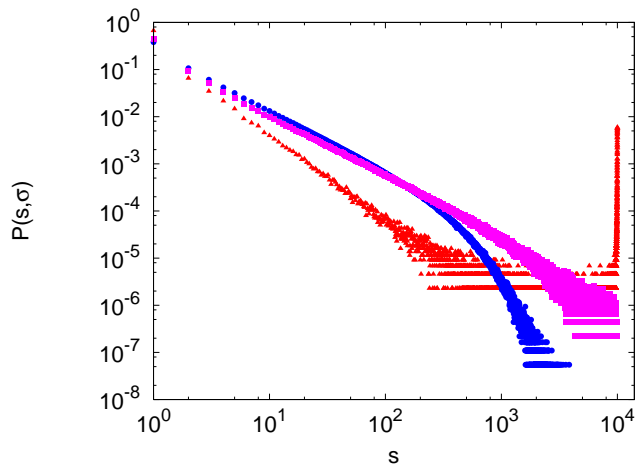} 
\caption{(Color online) Plot of $P(s,\sigma)$, the probability of an 
avalanche of size $s$ on a $100\times100$ triangular lattice for 
$\sigma=1.25$ (red triangles, lower curve with a peak at $s\approx 
10^4$), $\sigma=1.63$ (pink squares, curve with its tail nearest to 
the red triangles peak), and $\sigma=2$ (blue circles, curve with is 
tail farthest from the red/triangles' peak).} \label{fig2} 
\end{figure}

\begin{figure}[htb] 
\includegraphics[width=.5\textwidth,angle=0]{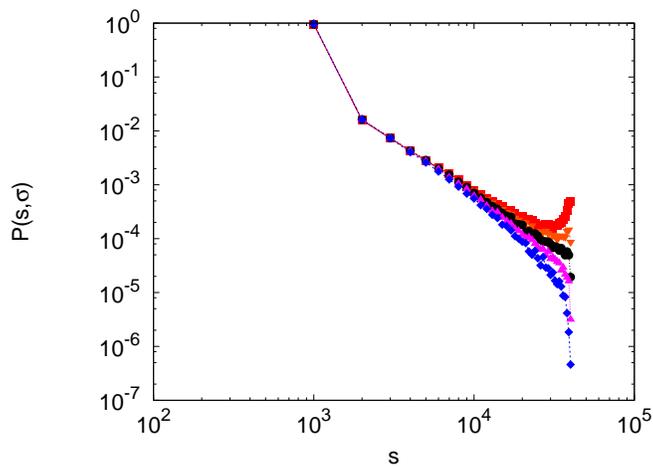} 
\caption{(Color online) Binned data for $P(s,\sigma)$ vs $s$ on a 
$200\times200$ triangular lattice for $\sigma=1.45$ (red squares), 
1.48 (orange inverted triangles), 1.50 (black circles), 1.52 (pink 
triangles), and 1.54 (blue diamonds). Generally, the last point in 
each curve may be ignored because of less data in the last bin. We 
estimate $\sigma_c=1.50$ for a lattice of linear size $L=200$ because 
the corresponding distribution is nearly linear over the entire range 
of avalanche sizes. Avalanches for $\sigma<\sigma_c$ tend to show a 
$\delta$-function peak at the largest avalanche, while avalanches for 
$\sigma>\sigma_c$ tend to bend down. The opposite trends for 
$\sigma<\sigma_c$ and $\sigma>\sigma_c$ become more pronounced as one 
moves farther away from $\sigma_c$. } \label{fig3} \end{figure}

\begin{figure}[htb] 
\includegraphics[width=.5\textwidth,angle=0]{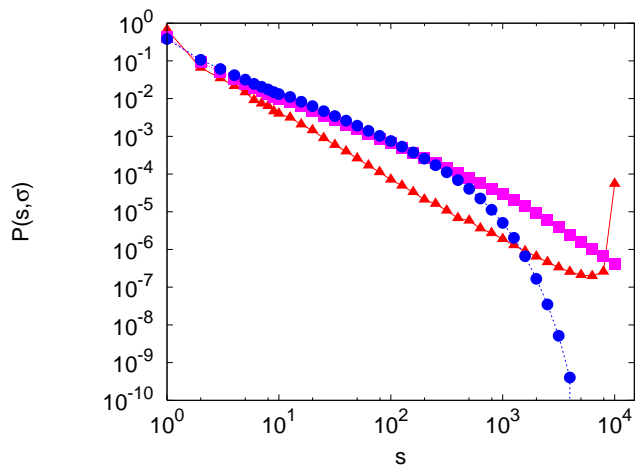} 
\caption{(Color online) Plot of $P(s,\sigma)$ using the same data as 
in Fig. 2 but using logarithmic binning. The colors and symbols have 
the same meaning as in Fig. 2.} \label{fig4} \end{figure}

\begin{figure}[htb] 
\includegraphics[width=.5\textwidth,angle=0]{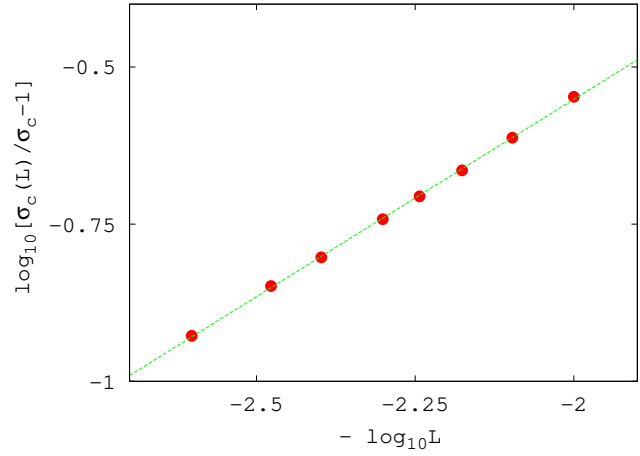} 
\caption{(Color online) Finite-size scaling: plot of - $\log_{10} L$ 
($x$ axis) vs $\log_{10} [\sigma_c(L)$/$\sigma_c$-1] ($y$ axis). The 
parameter $\sigma_c=1.27$ is determined by the best linear fit to the 
data of Table I. The slope of the line yields $\nu=1.6$.} 
\label{fig5} \end{figure}

\begin{figure}[htb] 
\includegraphics[width=.5\textwidth,angle=0]{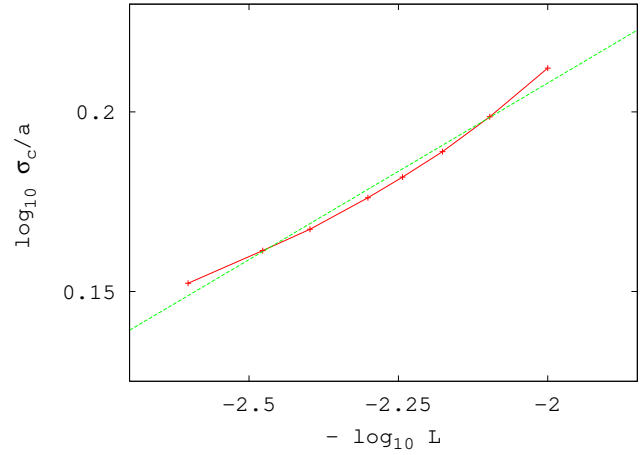} 
\caption{(Color online) Plot of - $\log_{10} L$ ($x$ axis) vs $ 
\log_{10} 
\sigma_c(L)/a $ for $a=1$ ($y$ axis). The curve does not fit well to 
a straight line. The straight line with a slope of 0.098 
($\nu=10.17$) shown in the figure is the best fit to the data. 
Changing the parameter $a$ shifts the curve along the $y$ axis.} 
\label{fig6} \end{figure}

\begin{figure}[htb] 
\includegraphics[width=.6\textwidth,angle=0]{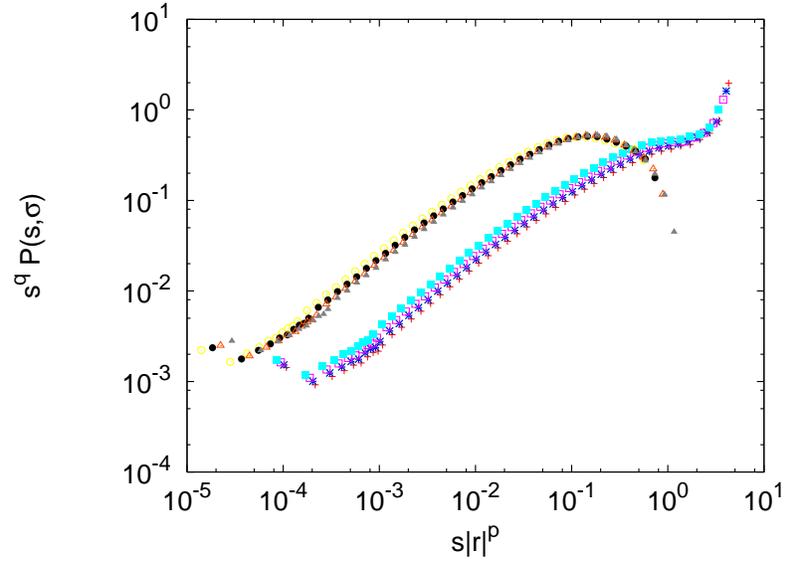} 
\caption{(Color online) Data collapse: plot of $s^q P(s,\sigma)$ vs 
$s|r|^p$ for a $200 \times 200$ lattice. The upper set of curves are 
for $\sigma=1.52, 1.53, 1.54, 1.56$ ($\sigma>\sigma_c=1.50$). These 
collapse reasonably on top of each other for $q=2$ and $p=0.66$. The 
lower set of curves are for $\sigma=1.48, 1.47, 1.46, 1.45$ 
($\sigma<\sigma_c$). These too show reasonable collapse for $q=2$ and 
$p=0.25$.} \label{fig7} \end{figure}

\end{document}